\documentclass[reprint,superscriptaddress,showpacs,preprintnumbers,amsmath,
amssymb,aps,10pt,twocolumn]{revtex4-1}
\usepackage{graphicx} 
\usepackage{natbib}

\usepackage[usenames,dvipsnames]{color}
\usepackage{graphicx,textcase} 
\usepackage{dcolumn,overpic} 
\usepackage[T1]{fontenc}
\usepackage{ae,aecompl}
\usepackage{chngcntr}
\usepackage[para]{threeparttable}
\usepackage{etoolbox}
\usepackage{lipsum}
\usepackage[bottom]{footmisc}
\usepackage{setspace}
\AtEndEnvironment{table*}{\vskip10pt}{}{}

\usepackage{url}
\usepackage{hyperref}
\hypersetup{colorlinks=true,breaklinks,linkcolor=blue,urlcolor=blue,citecolor=blue}

\begin{document}

\title{Half-metallicity and magnetism in the Co$_2$MnAl/CoMnVAl heterostructure}

\author{Igor Di Marco}
\affiliation{Department of Physics and Astronomy, Uppsala University, Box 516, SE-75120 Uppsala, Sweden}
\author{Andreas Held}
\affiliation{Theoretical Physics III, Center for Electronic Correlations and Magnetism, Institute of Physics, University of
Augsburg, D-86135 Augsburg, Germany}
\author{Samara Keshavarz}
\affiliation{Department of Physics and Astronomy, Uppsala University, Box 516, SE-75120 Uppsala, Sweden}
\author{Yaroslav O. Kvashnin}
\affiliation{Department of Physics and Astronomy, Uppsala University, Box 516, SE-75120 Uppsala, Sweden}
\author{Liviu Chioncel}
\affiliation{Theoretical Physics III, Center for Electronic Correlations and Magnetism, Institute of Physics, University of
Augsburg, D-86135 Augsburg, Germany}
\affiliation{Augsburg Center for Innovative Technologies, University of Augsburg, D-86135 Augsburg, Germany}

\date{\today}

\begin{abstract}
We present a study of the electronic structure and magnetism of Co$_2$MnAl, CoMnVAl and their heterostructure. We employ a combination of density-functional theory and dynamical mean-field theory (DFT+DMFT). We find that Co$_2$MnAl is a half-metallic ferromagnet, whose electronic and magnetic properties are not drastically changed by strong electronic correlations, static or dynamic. Non-quasiparticle states are shown to appear in the minority spin gap without affecting the spin-polarization at the Fermi level predicted by standard DFT. We find that CoMnVAl is a semiconductor or a semi-metal, depending on the employed computational approach.  We then focus on the electronic and magnetic properties of the Co$_2$MnAl/CoMnVAl heterostructure, predicted by previous first principle calculations as a possible candidate for spin-injecting devices. We find that two interfaces, Co-Co/V-Al and Co-Mn/Mn-Al, preserve the half-metallic character, with and without including electronic correlations. We also analyse the magnetic exchange interactions in the bulk and at the interfaces. At the Co-Mn/Mn-Al interface, competing magnetic interactions are likely to favor the formation of a non-collinear magnetic order, which is detrimental for the spin-polarization.
\end{abstract}

\maketitle 

\section{Introduction}
In the last decade spintronics has emerged as an important field of 
research at the intersection between electronics and magnetism~\cite{aw.fl.07}. 
The basic concept in spintronics is that information is transmitted 
by manipulating not only charge but also spin currents. 
Applications include various devices, as e.g. the magneto-resistive 
random access memory, based on the giant magnetoresistance (GMR) and
tunneling magnetoresistance (TMR) phenomena~\cite{ba.br.88,bi.gr.89}.
Practical realizations of these effects consist of multi-layered 
systems whose basic components are 
half-metallic ferromagnets (HMFs)~\cite{gr.mu.83,groo.91,gr.kr.86,wi.ha.89}
sandwiching a non-magnetic layer (either semiconducting or metallic).
In order to realize an efficient spin-injection, the HMFs have to be
chosen with a high spin-polarization, possibly holding over a wide
range of temperatures. 

Searching for materials with the right properties for spintronics 
is not a simple task. While experimental studies involve a 
substantial amount of resources, theoretical investigations suffer
of oversimplified modeling, neglecting important factors such as 
the non-stochiometry of the samples or the presence of 
defects~\cite{fa.ra.13}. Additionally, theoretical studies are
often based on density-functional theory (DFT), which
treats electronic correlations only approximately in
local or semi-local functionals~\cite{ho.ko.64,ko.sh.65,kohn.99}.

HMFs possess $3d$ electrons that are partially localized and are
therefore expected to exhibit significant correlation 
effects~\cite{ka.ir.08}. In fact they are characterized by
an essential feature
due to many-body effects, i.e. the existence of non-quasiparticle
(NQP) states~\cite{ed.he.73,ir.ka.90,ir.ka.94}. NQP states influence
the value and temperature dependence of the spin polarization in 
HMFs~\cite{ch.sa.08,ir.ka.94,ka.ir.08} and are therefore of primary
interest for potential applications. NQP states have been
shown to contribute significantly to the tunneling transport 
in heterostructures containing 
HMFs~\cite{ir.ka.02,mc.fa.03,tk.mc.01,mc.fa.02,ir.ka.06},
even in presence of arbitrary disorder. 
The origin of NQP states is connected with ``spin-polaron''
processes; in half-metals where the gap appears in the
minority spin channel (minority spin gap half-metals)
no low-energy electron excitations are possible in 
the minority spin channel within the single-particle 
picture. However, low-energy electron excitations can still be
constructed as superpositions of spin-up electron excitations 
and virtual magnons~\cite{ed.he.73,ir.ka.90,ir.ka.94,ka.ir.08}. 

Including NQP states in the electronic structure requires methods
beyond standard DFT. The dynamical mean-field theory
(DMFT)~\cite{me.vo.89,ge.ko.96,ko.vo.04} describes the local 
correlation effects exactly and is therefore suitable
to capture the essential physics of the HMFs~\cite{ka.ir.08}.
To retain the predictive character of ab-initio calculations
one often employs a combination of DFT in the local spin-density
approximation (LSDA) and DMFT, which we address here with
the acronym LSDA+DMFT (for a review of this approach, see 
Ref.~\onlinecite{ko.sa.06}). The LSDA+DMFT scheme has
been successfully applied to describe the physical properties
of a variety of HMFs, including systems of both minority
spin gap~\cite{ch.ka.03,ch.ka.05,ch.ar.06,gr.mu.83,ch.al.07,ch.ke.16}
and majority spin gap~\cite{ch.ar.06,ch.ar.09}. The general
feature reported in these studies is that NQP states leave their signature
on the excitation spectra as small resonances just above (below) 
the Fermi level when the gap lies in the minority (majority) 
spin channel. Non-local correlations have also been investigated,
by means of theories beyond DMFT, as e.g. the variational cluster 
approach~\cite{po.ai.03,pott.03.se,ai.ar.05,ai.ar.06}. These 
calculations not only confirm the main findings obtained with
DMFT, but they even assign a larger weight
to the NQP states around the 
Fermi level~\cite{ch.al.07,al.ch.08.sp,al.ch.10}. 

%
%

While the depolarizing mechanism associated to NQP states has been 
widely studied in bulk HMFs, a few studies analysed this phenomenon
in multilayered heterostructures which are the components of the
GMR/TMR setup. The magnetic tunnel junction Co$_2$MnSi/AlO$_2$/Co$_2$MnSi
was investigated by means of tunneling spectroscopy 
measurements, showing evidence of NQP states above the Fermi 
level~\cite{ch.sa.08}. More recently, the magnetic 
tunnel junction Co$_2$MnSi/MgO was analysed through extremely
low energy photoemission spectroscopy~\cite{fe.st.15}. 
Although these results can be interpreted in terms of NQP 
states, the authors suggested that they can also be a 
signature of a non-collinear arrangement of spins at the 
interface. These experimental findings and the lack of 
material-specific studies of interfaces based on advanced 
many-body theories are the motivations behind the present work.
To analyse the role of NQP states and the tendency to
non-collinear magnetism, we focus on 
bulk Co$_2$MnAl and CoMnVAl, as well as their 
heterostructure. The relevance of these systems 
lays in the fact that they were previously 
predicted~\cite{ch.gr.11} to form two
interfaces preserving the half-metallic
character of the parent material Co$_2$MnAl, namely
Co-Co/V-Al and Co-Mn/Mn-Al. In the present work we show 
that this important conclusion is not changed by the
inclusion of strong correlation effects. We also show
that the Co-Mn/Mn-Al interface is characterized by competing
exchange interactions that may potentially lead to a non-collinear 
arrangement of spins. Our results suggest that the present
heterostructure is a good candidate for the 
experimental realization of a highly spin-polarized interface,
in particular if focusing on Co-Co/V-Al.

The paper is organized as follows. Sec.~\ref{sec:method} illustrates
the Computational Details used for our calculations. Sec.~\ref{sec:co2mnal}
and Sec.~\ref{sec:comnval} are dedicated to bulk Co$_2$MnAl and 
bulk CoMnVAl, respectively. The magnetic and spectral properties of the
heterostructure composed by Co$_2$MnAl and CoMnVAl are presented 
in Sec.~\ref{sec:interface}. The Conclusions of our work are 
drawn in Sec.~\ref{sec:conclusions}.

\section{Computational details}
\label{sec:method}
Scalar relativistic electronic structure calculations were performed 
by means of the full-potential linearized muffin-tin
orbitals (FP-LMTO) code RSPt~\cite{rspt_book,rspt_website}. 
The exchange-correlation potential was treated in the LSDA and 
in the generalized gradient approximation (GGA), using respectively the 
parametrizations by Perdew and Wang~\cite{pe.wa.92} and by
Perdew, Burke and Enzerhof~\cite{perdew96prl77:3865}. 
The muffin-tin spheres, 
which in a full-potential code are still used
to divide the physical space in the unit cell, were carefully
optimized to offer a good description of the electron density
in both bulk and heterostructure. This procedure lead to 
muffin-tin radii of 2.20, 2.10, 1.8 and 2.34 a.u. for 
respectively Mn, Co, Al and V. The valence electrons were
described with $4s$, $4p$ and $3d$ states for the transition 
metals and $3s$, $3p$ and $3d$ states for Al. Selected
calculations were also performed with an extended set of 
valence electrons, including semi-core states. Considering
these additional states resulted into a variation of the 
equilibrium lattice constants of less than 0.1\% and was
therefore deemed not necessary for the purposes of this
study. 

In RSPt one can address the effects of strong Coulomb
interaction for localized electrons at the level of 
LSDA+U~\cite{an.ar.97l} and LSDA+DMFT~\cite{ko.sa.06}. 
Both approaches
start from a correction to the Kohn-Sham Hamiltonian in the form of 
an explicit intra-atomic Hubbard repulsion for the $3d$ states
of the transition metal elements. 
This term can then be 
treated self-consistently in a mean-field (Hartree-Fock) approach, 
as in LSDA+U~\cite{an.ar.97l}, or by more sophisticated many-body 
theories, as in LSDA+DMFT~\cite{ko.sa.06}. We here consider the most
general form of LSDA+U and LSDA+DMFT, where the interaction vertex 
is taken as a full spin and orbital rotationally invariant
4-index U-matrix~\cite{an.ar.97l,sh.dr.05}. This matrix is parametrized
in terms of Slater parameters, which are $F^0$, $F^2$ 
and $F^4$ for $3d$ states. These parameters can in turn be obtained from
the average screened Coulomb interaction $U$ and corresponding Hund exchange 
$J$~\cite{gr.ma.07,ko.sa.06}. Calculated Coulomb interaction parameters
were recently reported for Co$_2$MnAl~\cite{sa.ga.13}, based
on the constrained random-phase approximation (cRPA). The authors
reported a partially (fully) screened value of $U$ of 
3.40 (0.81) eV for Co 
and of 3.23 (0.83) eV for Mn. As far as we know, there are no
calculations available for CoMnVAl. Therefore, for simplicity 
and for facilitating a meaningful comparison between bulk and 
heterostructures, we here adopted a uniform value of $U=2.0$ eV
for the $3d$ states of all transition metals (Mn, Co and V).
This value is included in the range defined by the partially
screened value and the fully screened value reported above,
and is also in line with the values calculated for the corresponding
elemental solids~\cite{ar.ka.06,mi.ar.08}. The Hund's exchange
$J$ is not affected much by screening and was here set to a 
value of 0.9 eV. For completeness, we also performed an 
additional set of calculations by using the partially 
screened values mentioned above. Those results are 
summarized in Appendix~\ref{sec:calculatedU}.

The addition of a Hubbard U interaction term in the total energy functional,
also introduces the need for a ``double-counting'' correction. The latter
accounts for the fact that the contribution to the total energy 
due to the Hubbard term is already included (although not correctly) 
in the exchange correlation functional. 
The double-counting scheme is unfortunately not uniquely defined, 
and usually creates some ambiguity~\cite{pe.ma.03,yl.pi.09}. 
Here we consider the double counting in the so-called fully localized
limit (FLL)~\cite{an.ar.97l,pe.ma.03,li.an.95} in LSDA+U. For LSDA+DMFT
we instead remove the orbital average of the static part of the 
self-energy~\cite{ka.li.02,gr.ma.07,ma.mi.09}. The effective impurity model 
arising in LSDA+DMFT is solved by means of the relativistic 
version of the Spin-Polarized T-Matrix Fluctuation Exchange (SPTF)
approximation~\cite{ka.li.02,po.ka.05}. The applicability of SPTF
to HMFs and Heusler compounds has already been demonstrated in a
number of studies, as discussed in a recent review~\cite{ka.ir.08}.

The LSDA+DMFT calculations were performed using finite temperature
Green's functions at $T=160$ K. A total of 2048 Matsubara frequencies
were considered. Spectral functions for all methods were calculated 
at real energies displaced at a distance $\delta=2$ mRy from the real 
axis. In case of LSDA+DMFT the analytical continuation of the
self-energy was required to obtain the spectral function. To this aim
we used the least-square average Pad\'e approximant 
method~\cite{sc.lo.16}, whose initial steps are based on Beach's 
formulation~\cite{be.go.00}. In the following, for an 
easier comparison with electronic structure literature, we will
drop the term spectral function in favour of density of states (DOS) 
and projected density of states (PDOS). The local orbitals
used to describe the $3d$ states in LSDA+U and LSDA+DMFT were
constructed from the so-called muffin-tin (MT) heads and are 
therefore atomic-like. A more comprehensive discussion on this and
other technical details can be found in  
Refs.~\onlinecite{gr.ma.07,ma.mi.09,gr.ma.12}.

We have also calculated the exchange parameters between the magnetic atoms in the bulk and at the interface of the two compounds. 
Using the converged electronic density obtained in the previous step, the interatomic exchange parameters were obtained by mapping the magnetic excitations onto the Heisenberg Hamiltonian:
\begin{equation}
\hat{H}=-\sum_{i \neq j}J_{ij}\vec{e}_i\vec{e}_j
\end{equation}
where $J_{ij}$ is an exchange interaction between the spins at sites $i$ and $j$, and $\vec{e}_i$ is a unit vector along the magnetization direction at the corresponding site. We extracted the $J_{ij}$'s by employing the method of infinitesimal rotations of the spins in the framework of the magnetic force theorem~\cite{lich-jij-1,lich-jij-2}. More technical details about the evaluation of the exchange interactions in RSPt can be found in Ref.~\cite{jij}. The local orbitals used to calculate the $J_{ij}$'s were again chosen as MT-heads.

While we used RSPt for all data presented in this article, the
geometry optimization of the heterostructure was too heavy to
be performed in an all-electron method. Therefore, to this 
aim, we employed the projector augmented wave method~\cite{paw}, 
as implemented in the VASP code~\cite{vasp}. The plane-wave energy 
cut-off was set to 550 eV to ensure the completeness of the basis
set. The positions of the ions as well as the volume were relaxed 
so that the residual forces between the ions became less that 
0.01 eV/\AA~. Meanwhile, the lattice constant in the \textit{xy}
plane as well as in the \textit{z} direction have been varied 
separately around the averaged of the optimized bulk values for
Co$_2$MnAl and CoMnVAl, until the change in the total energy was
found converged up to 0.001 eV.

\section{bulk}
\subsection{The half-metallic Co$_2$MnAl}
\label{sec:co2mnal}
Co$_2$MnAl~\cite{bu.en.81} is a type of  Heusler compound which crystallizes  
in the cubic $L2_1$ structure (space group $Fm\overline{3}m$). The Co atoms
are placed  on  Wyckoff  position $8c /  (1/4,1/4,1/4)$, while Mn  and  Al  
atoms are situated in the position $4a (0,0,0)$  and  $4b (1/2,1/2,1/2)$ 
respectively. We performed relaxation of the lattice constant in GGA,
which usually leads to a good description of the chemical bonding
in transition metals compounds. The Brillouin Zone (BZ) was sampled
with a dense Monkhorst-Pack grid of 16 $\times$ 16 $\times$ 16 
$\mathbf{k}$-points. We obtained an equilibrium lattice constant
of $5.69$ \AA, which coincides with the value reported in a recent
study based on a pseudo-potential plane-wave code~\cite{zh.ji.15}. 
This value is in reasonable agreement with the experimental value of
$5.75$ \AA~\cite{bu.en.81}, which confirms that our approach can
describe the structural properties well.

\begin{figure}[t]
\centering
\includegraphics[width=0.85\linewidth]{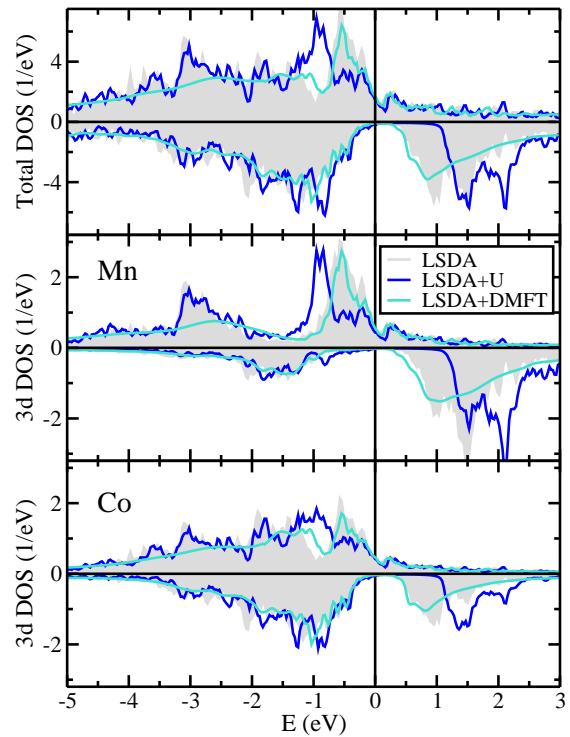}
\caption{(color online) Total DOS of Co$_2$MnAl, as well as PDOS for 
the $3d$ states of Mn and Co, as obtained in LSDA (gray shade), 
LSDA+U (blue) and LSDA+DMFT (turquoise). Positive and negative values 
correspond to majority and minority spin channels, respectively.
The Fermi level is at zero energy and emphasized with a 
vertical line.}
\label{co2mnal_dos}
\end{figure}

We then performed an analysis of magnetic and spectral properties
at the equilibrium volume in LSDA, LSDA+U and LSDA+DMFT. The 
spin-resolved DOS and PDOS for the $3d$ states of Mn and Co 
in the three methods are reported in Fig.~\ref{co2mnal_dos}. In 
plain LSDA Co$_2$MnAl is a half-metal with a gap in the minority 
spin channel of about 0.6 eV, consistently with previous
literature~\cite{ku.fe.07,ji.ya.08}. The region of low-energy
excitations, around the Fermi level, is dominated by the $3d$ 
states of the transition metals.
The $3d$ states of two Co sublattices couple and form 
bonding hybrids, which also hybridize with the
manifold of $3d$ states of Mn, for both $e_{g}$ and $t_{2g}$ 
symmetries~\cite{ka.ir.08}. This coupling also results
in Co-Co antibonding hybrids, which remain uncoupled,
owing to their symmetry. Therefore the gap is 
defined by the triply degenerate Co-Co antibonding hybrids 
of $t_{2g}$ character and the doubly degenerate Co-Co
antibonding hybrids of $e_{g}$ character. Fig.~\ref{co2mnal_dos}
illustrates how the energy gap in the minority spin channel 
is defined by Co states, while Mn-derived states are 
characterized by a larger gap of about 1.5 eV.

When adding exchange and correlation effects at 
the LSDA+U level, one
observes an increase of the gap. In the majority spin
channel one observes that the large peak of $t_{2g}$ character
moves from -0.5 eV in LSDA to -0.9 eV in LSDA+U, while 
the smaller peaks of $e_{g}$ character remain pinned at
the Fermi level and at around -3 eV. In the minority 
spin channel, instead, both symmetries are affected
and the spectrum is almost uniformly shifted upwards
of 0.5 eV.

\begin{figure}[b]
\centering
\includegraphics[width=0.85\linewidth]{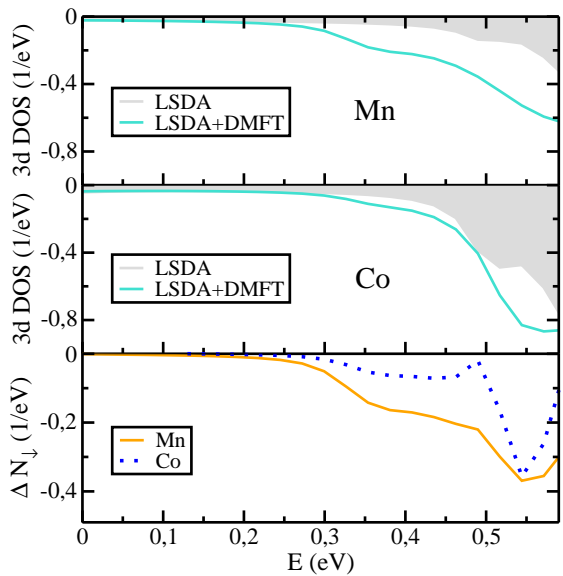}
\caption{(color online) Minority spin PDOS for the Mn-$3d$ states and
for the Co-$3d$ states in Co$_2$MnAl, as obtained in LSDA (gray shade) 
and LSDA+DMFT (turquoise). Negative values are used, to have the 
same convention as in Fig.~\ref{co2mnal_dos}.
The difference $\Delta N_\downarrow(E)$ 
between the PDOS obtained in LSDA and LSDA+DMFT for each set of 
states is also shown, in the bottom panel. 
The Fermi level is at zero energy.}
\label{co2mnal_inset}
\end{figure}
 
\begin{figure}[t]
\centering
\includegraphics[width=0.9\linewidth]{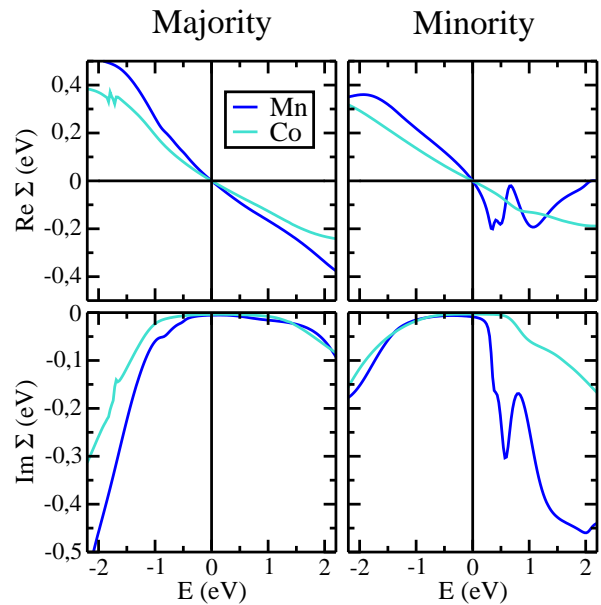}
\caption{(color online) Average self-energy per orbital of the Mn-$3d$ states 
and Co-$3d$ states in Co$_2$MnAl, separately for majority (left panels) and 
minority (right panels) spin channels. Both real and imaginary 
parts are reported, respectively in the top and bottom panels. The 
Fermi level is at zero energy and emphasized with a vertical line.}
\label{co2mnal_sigma}
\end{figure}

In LSDA+DMFT one obtains a physical picture closer 
to the one given by standard LSDA than to the one
given by LSDA+U. With respect to LSDA, two 
interesting effects are visible.
At high energies
the spectrum is smeared and shrinked towards the Fermi
level, in analogy to what happens in the itinerant 
ferromagnets Fe, Co and Ni~\cite{gr.ma.07}. Second, 
at low energies, NQP states appear in the minority spin 
channel, just above the Fermi energy. These states
lead to a smaller gap but do not destroy the 
half-metallic character. An interesting feature, which is 
just visible in Fig.~\ref{co2mnal_dos}, is that NQP states
affect mainly Mn-$3d$ states, while the effects at the Co sites 
is much reduced. This can be seen better in 
Fig.~\ref{co2mnal_inset}, where the minority spin PDOS curves 
shown in Fig.~\ref{co2mnal_dos} are enlarged in the closest 
region above the Fermi energy. 
A more quantitative measure of this asymmetry between Mn and 
Co can be seen in the bottom panel of Fig.~\ref{co2mnal_inset}, 
where the difference $\Delta N_\downarrow(E)$ between the minority
spin PDOS obtained in LSDA+DMFT and the one obtained in LSDA is reported.
The large negative peak at about 0.5 eV is due to
the shift of the band edge, but the shoulder at 0.3-0.4 eV
can be ascribed to NQP states. The effects on NQP states
on the spectrum at the Mn site are about 5 times larger
than those on the spectrum at the Co sites. The signature
of NQP states is particularly evident in the 
orbitally-averaged self-energy function, which 
is reported in Fig.~\ref{co2mnal_sigma}. 
In the minority spin channel, the imaginary part of 
the Mn self-energy is characterized by a large peak 
appearing at 0.4 eV above the Fermi level, while only
a small shoulder is visible for Co. This is in sharp 
contrast with the majority spin channel, where
curves have similar shapes. One can also notice that
the magnitude of the corrections induced by the 
self-energy for Mn is much bigger than for Co.
This is in 
agreement with studies on the transition metal elements,
where correlation effects were shown to be the larger,
the closer the $3d$ shell is to half-filling, i.e. 
the larger the magnetic moment~\cite{ma.mi.09}

\begin{figure}
\centering
   \includegraphics[width=0.9\linewidth]{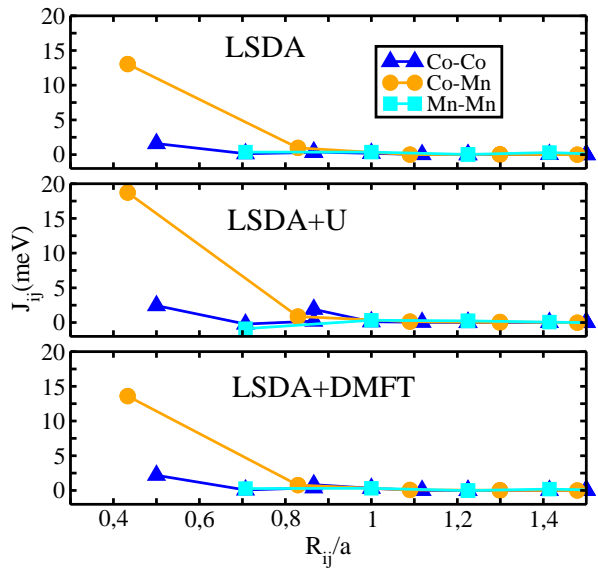}
\caption{Interatomic exchange parameters $J_{ij}$ between all magnetic atoms in bulk Co$_2$MnAl as given by LSDA, LSDA+U and LSDA+DMFT. Interatomic distances are expressed in units of the equilibrium lattice constant.}
\label{fig:j-bulk}
\end{figure}

\begin{table}[b]
\begin{tabular}{cccccc}
\hline
     &  $\mu_{Mn}$ &  $\mu_{Co}$ & $\mu_{Al}$ &  $\mu_{int}$ &  $\mu_{cell}$ \\
\hline
LSDA      & 2.48 & 0.81 & -0.03 & -0.07 & 4.00 \\
LSDA+U    & 2.63 & 0.79 & -0.04 & -0.17 & 4.01 \\
LSDA+DMFT & 2.46 & 0.82 & -0.03 & -0.08 & 4.00 \\
\hline
\end{tabular}
\caption{Site-projected and total magnetic moments for bulk Co$_2$MnAl, as computed
in LSDA, LSDA+U and LSDA+DMFT. The values are given in $\mu_{B}$.\label{tab:HMF_bulk_moments}}
\end{table}
Magnetic properties can be analysed through the total and site-projected
magnetic moments, which are reported in Table~\ref{tab:HMF_bulk_moments}, for LSDA, 
LSDA+U and LSDA+DMFT. While LSDA and LSDA+DMFT lead to similar values,
in LSDA+U one can see that the moment at the Mn site is increased,
which is compensated by an increase of the anti-parallel moment
in the interstitial region (in between the muffin-tin spheres).
This change is not determined by the redistribution of the total 
charge in between the two regions but by a genuine increase of the
exchange splitting. This remodulation is also 
accompanied by a smaller change of magnetic moment at the Co sites.

Using the electronic structure obtained in LSDA, LSDA+U and LSDA+DMFT,
we calculated interatomic exchange parameters $J_{ij}$'s. Results 
are shown in Fig.~\ref{fig:j-bulk}. In bulk Co$_2$MnAl, the $J_{ij}$'s 
decay quite fast with the interatomic distance and therefore the 
dominant coupling is the nearest-neighbour ferromagnetic interaction 
between Co and Mn atoms. The present feature 
is in good agreement with prior DFT studies~\cite{com-j-bulk} and 
is independent of the method used for treating correlation effects. 
Consistently with the PDOS (Fig.~\ref{co2mnal_dos}), the $J_{ij}$'s 
obtained from LSDA and LSDA+DMFT are quite similar to each other, whereas 
LSDA+$U$ results show larger differences. The dominant Co-Mn coupling is 
enhanced in LSDA+$U$, which reflects the increase of the magnetic moment
reported in Table~\ref{tab:HMF_bulk_moments}.

\subsection{The semiconducting CoMnVAl}
\label{sec:comnval}
CoMnVAl crystallizes in the Y structure (LiMgPdSn type,
{$F\bar{4}3m$} symmetry; see e.g. 
Refs.~\onlinecite{we.zi.88}), where the Wyckoff
positions $4a (0,0,0)$, $4b (1/2,1/2,1/2)$,
$4c (1/4,1/4,1/4)$ and $4d (3/4,3/4,3/4)$ are respectively
occupied by Mn, Co, Al and V. It can be synthesised
as a solid solution by mixing Mn$_2$VAl and Co$_2$VAl
in equal amount~\cite{ba.fe.11}.
The primitive cell contains 24 valence electrons and,
according to the Slater-Pauling 
rule~\cite{slat.36,paul.38}, should be expected
to be a semiconductor (SCs), either magnetic or
non-magnetic~\cite{ta.sa.16}. Among the quaternary Heusler
compounds, one is more likely to find small-gap SCs, since
the unit cell contains three different transition 
metals~\cite{oz.sa.13}. CoMnVAl was in fact predicted to be
a non-magnetic SC through first principles 
calculations~\cite{ch.gr.11,oz.sa.13}, although
the presence of an indirect negative gap formally makes 
it a semi-metal~\cite{ch.gr.11,ta.sa.16}.
The lattice constant measured
in experiments amounts to 5.80 \AA~\cite{ba.fe.11}, which is 
rather close to that of Co$_2$MnAl. These characteristics make
CoMnVAl and Co$_2$MnAl suitable to be grown in heterostructures 
useful for spintronics~\cite{ch.gr.11}.

As for Co$_2$MnAl, we performed relaxation of the lattice constant
in GGA. The BZ was again sampled
with a dense Monkhorst-Pack grid of 16 $\times$ 16 $\times$ 16 
$\mathbf{k}$-points. We obtained an equilibrium lattice constant
of $5.74$ \AA, which is in line with the values obtained with 
a pseudo-potential plane-wave method~\cite{wa.da.15}
and with a full-potential linearized augmented plane-wave (FP-LAPW) 
method~\cite{ta.sa.16}, respectively $5.76$ {\AA} and $5.74$ {\AA}.
The calculated lattice constant is in reasonable agreement with the
experimental value reported above, leading to an error of about 1\%.

\begin{figure}
\centering
\includegraphics[width=0.8\linewidth]{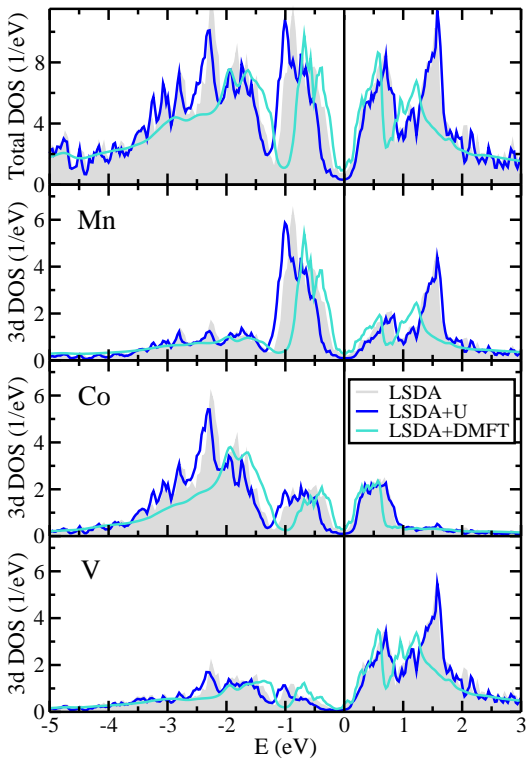}
\caption{(color online) Total DOS of CoMnVAl, as well as PDOS for 
the $3d$ states of Mn, Co and V, as obtained in LSDA (gray shade), 
LSDA+U (blue) and LSDA+DMFT (turquoise). Only the total trace is shown,
since the system is non-magnetic. 
The Fermi level is at zero energy and emphasized with a 
vertical line.}
\label{dossc}
\end{figure}

\begin{figure}[t]
\centering
   \includegraphics[width=0.9\linewidth]{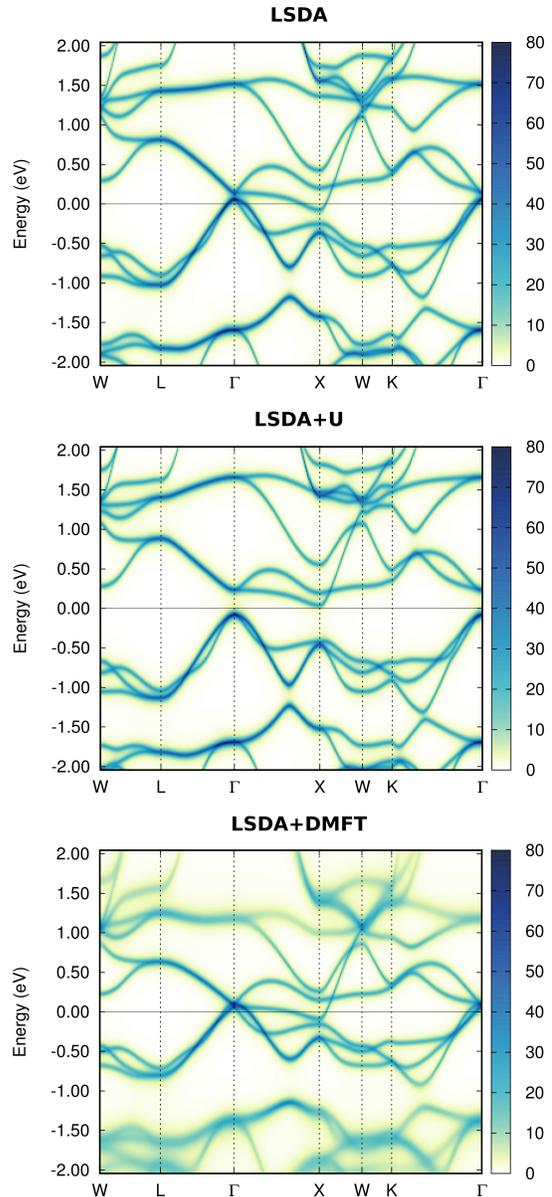}
\caption{Total spectral densities of CoMnVAl along high-symmetry directions 
in the Brillouin zone, as obtained in LSDA, LSDA+U and LSDA+DMFT.
The Fermi level is at zero energy.}
\label{fig:bandsc}
\end{figure}

In the ground-state there are no finite magnetic moments,
consistently with previous literature. The spin-integrated
DOS and PDOS of CoMnVAl are reported in 
Fig.~\ref{dossc}, for LSDA, LSDA+U and LSDA+DMFT. The data
obtained in LSDA show a small amount of states at the Fermi
energy. 
The origin of these states is clarified by inspecting
the LSDA band structure, reported in the top panel 
of Fig.~\ref{fig:bandsc}. The valence band crosses 
the Fermi level at the $\Gamma$-point, while the 
conduction band goes below it at the $X$-point. 
These semi-metallic features have been reported in prior works~\cite{ch.gr.11,ta.sa.16}.
Further insight into the band structure can be obtained by analysing
projections over real spherical (cubic) harmonics (data not shown).
One can identify that the contribution at the $\Gamma$ point arises
from Mn and Co $t_{2g}$ states, while at the $X$ point there are 
mainly V-$e_g$ states. The latter hybridize with Mn and Co $e_g$
orbitals to form the conduction band extending from $X$ to $\Gamma$.

In LSDA+U CoMnVAl turns into a proper SC. This is not fully visible 
in Fig.~\ref{dossc}, due to the smearing used for printing DOS and 
PDOS (see Computational Details). Nevertheless, the band structure 
reported in the middle panel of Fig.~\ref{fig:bandsc}
shows this feature unambiguously. The formation of an indirect 
band gap along the $\Gamma$-$X$ direction is due to the fact
that the V-$e_g$ states (together with the corresponding hybridizing
states of Co and Mn) are pushed upwards by the Hubbard U 
correction and become completely unoccupied.

The physical picture emerging from LSDA+DMFT calculations is more complex.
The mechanism which leads to the opening of a band gap in LSDA+U is 
in fact counteracted by the fact that the occupied $3d$ states are pushed
upwards. This is clear in both Fig.~\ref{dossc} and in the bottom panel of
Fig.~\ref{fig:bandsc}, especially around the $\Gamma$ point. 
As a consequence, the top of the valence band, corresponding to Mn and Co
$t_{2g}$ states, almost touches the conduction band, corresponding to Mn 
and Co $e_{g}$ states. In the bottom panel of Fig.~\ref{fig:bandsc}
one can also observe that at high excitation energies
the band structure is no longer well defined, due to shorter 
quasiparticle lifetimes arising from electron-electron interaction.

A more quantitative analysis of the changes induced by the three
computational methods can be made by extracting the $\Gamma \to X$ 
band gap from Fig.~\ref{fig:bandsc}. In LSDA we obtain
a value of -0.14 eV, which is in fair agreement with the value of 
-0.07 eV obtained in GGA with a FP-LAPW code~\cite{ta.sa.16}. In 
LSDA+DMFT, instead, we obtain a value of -0.21 eV, which corresponds
to a variation of around 0.07 eV with respect to the LSDA value. This
change is identical to the one obtained in GW with respect to 
standard GGA, as presented in Ref.~\onlinecite{ta.sa.16}. 
GW data  are also similar to our LSDA+DMFT results in the reduction 
of the direct band gap at $X$ point, with respect to standard DFT.
Conversely the direct band gap at the $\Gamma$ point is increased
in GW, while almost vanishes in our LSDA+DMFT simulations. Since
the impurity solver used here (SPTF) contains diagrammatic 
contributions that are similar to those included in GW, we 
can attribute the observed discrepancies to the fact that 
DMFT neglects 
non-local fluctuations, while they are included in GW. Overall,
however, our data agrees well with previous literature.

In the following, for simplicity, we will refer to CoMnVAl as a SC, 
although it is such only in LSDA+U.

\begin{figure}[t]
\centering
\includegraphics[width=1.0\linewidth]{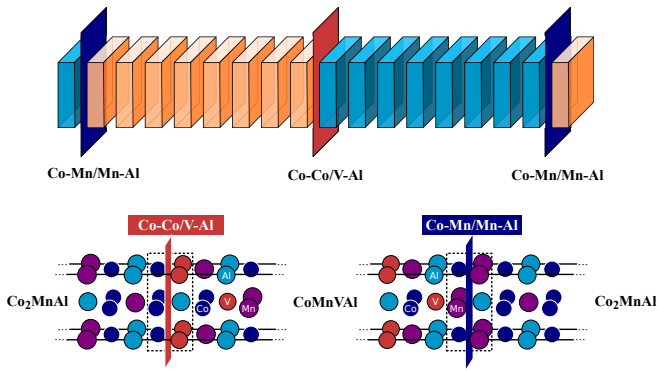}
\caption{(color online) Supercell composed by 8 unit cells of the HMF
Co$_2$MnAl (orange blocks) and by 8 unit cells of the SC CoMnVAl (turquoise
blocks). The supercell contains two distinct interfaces, namely
Co-Co/V-Al (red plane) and Co-Mn/Mn-Al (blue panel), whose composition
is shown in the lower panels. The atoms of Mn, Co, Al and V are 
respectively represented by violet, blue, turquoise and red spheres.}
\label{fig:interface}
\end{figure}

\begin{figure*}[th!]
\centering
\includegraphics[width=0.9\linewidth]{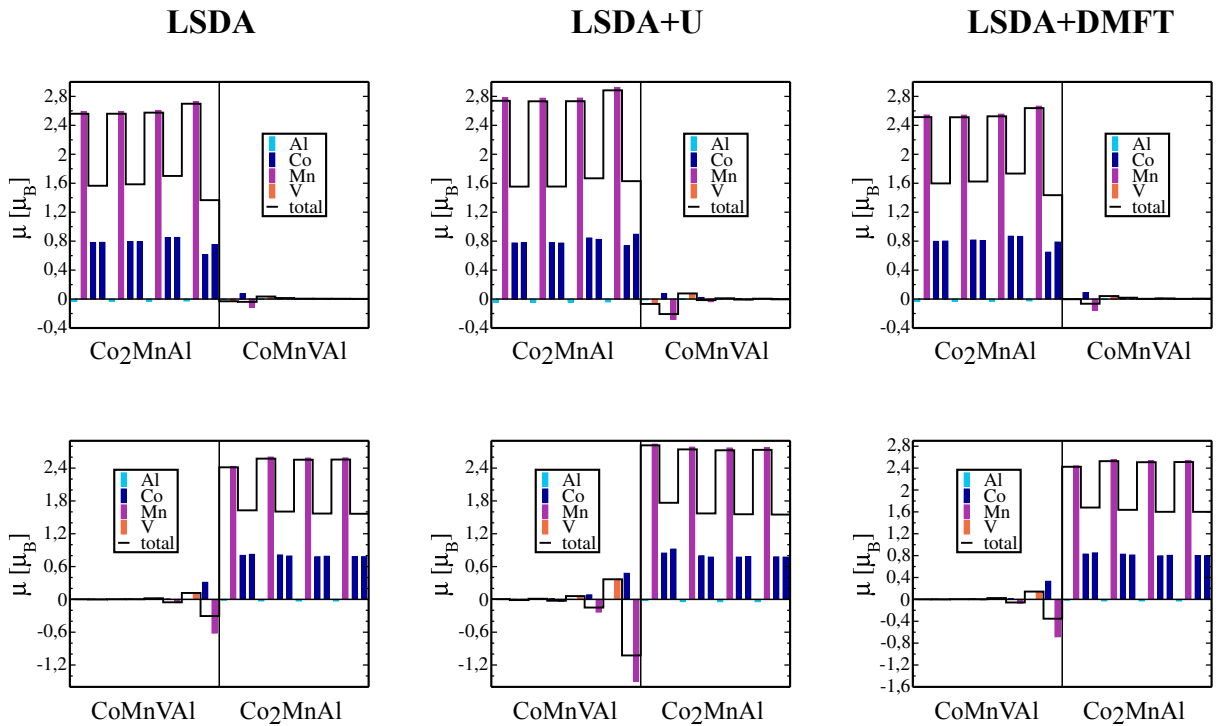}
\caption{(color online) Trends of the magnetic moments across the interface in LSDA,
LSDA+U and LSDA+DMFT. The top panels refer to the Co-Co/V-Al interface, while the
bottom panels refer to the Co-Mn/Mn-Al interface. The $x$ axis indicates the 
different layers along the (001) direction. The moment per layer is given
by the thick black line, while the coloured bars indicate the moment at 
each atomic site within a layer. The colours are set using the same convention
of the atomic spheres of Fig.~\ref{fig:interface}.}
\label{fig:moments}
\end{figure*}

\section{Half-metallic interface}
\label{sec:interface}
As discussed in the Introduction, HMFs are usually incorporated in 
spintronic devices in the form of films or multilayers and 
therefore it is important to address the properties of these systems.
De Wijs and de Groot~\cite{wi.gr.01} were the first authors to 
investigate surfaces and interfaces of HMFs in search for good 
candidates for spin-injection. Although they found out that 
surfaces and interfaces of HMFs are by construction prone 
to a loss of spin-polarization, they also identified 
NiMnSb/CdS (111) as a system preserving the half-metallicity
of its parent material. Later, various studies addressed the 
problem of interfacing HMFs with different types of SCs 
in the search of good heterostructures for 
spintronics~\cite{ma.ga.04b,pi.co.03,ma.le.05,ha.kr.05,na.mi.06,ko.ha.09}.
The presence of interface states reducing (or even inverting)
the spin polarization has been explained as due to the lack of
the d-d hybridization at the interface, which leaves
d-like dangling bond states inside the gap~\cite{ga.ma.06}.
To solve this problem, Chadov {\itshape{et al.}}~\cite{ch.gr.11} 
suggested that Heusler alloys have such a rich variety of 
physical properties that one may attempt to construct interfaces
where both the HMF and the SC are Heusler alloys. This 
would make it possible to preserve the nature of the 
bonding through the interface, which
may potentially preserve the complete spin-polarization.
Co$_2$MnAl and CoMnVAl were chosen as the two fundamental
components for the interface. Among all possible stackings,
Co-Co/V-Al and Co-Mn/Mn-Al were identified as
energetically favourable interfaces which preserve 
the half-metallic character,
and therefore suitable for experimental
realisation~\cite{ch.gr.11}. 

In this section we aim to explain under what conditions 
the coherence of the bonding through the interface
remains persistent against the effects of electron correlations.
We performed electronic structure calculations
of the heterostructure composed by Co$_2$MnAl and CoMnVAl.
We considered a supercell of 64 atoms (32 atoms
for each component) aligned along the (001) direction,
as illustrated in Fig.~\ref{fig:interface}. The 
structure was first fully relaxed in GGA 
by means of the VASP code, as described above. The
BZ was sampled with a dense Monkhorst-Pack grid of 
8 $\times$ 8 $\times$ 2 $\mathbf{k}$-points, which
lead to total energies converged up to 0.1 meV. 
The equilibrium lattice constant of the heterostructure is
of about 5.718 \AA, which is roughly the average of the
optimized bulk values reported above, and is in good
agreement with the results by
Chadov {\itshape{et al.}}~\cite{ch.gr.11}. The relaxed 
structure was used as an input for electronic structure
calculations with RSPt. For those, the BZ was sampled 
with a dense Monkhorst-Pack grid of 
8 $\times$ 8 $\times$ 1 $\mathbf{k}$-points.

\subsection{Magnetic properties}
First, we focus on the magnetic moments per atomic site in LSDA, 
which are reported in
Fig.~\ref{fig:moments}. For convenience, also the magnetic moments
per layer are shown (thick black line). At the Co-Co/V-Al interface, 
the symmetry of the Co sites is broken as they acquire two different 
magnetic moments. These moments are slightly smaller than those
at the Co sites in the innermost layer, which in turn reproduces
the bulk properties quite well (see Appendix~\ref{sec:supercell}).
On the SC side of the interface, the exchange coupling with the
Co-$3d$ states induces a small moment at the V site. In the 
next layer, the moments on Co and Mn align anti-parallel to each other, 
leading to a small total moment. The non-magnetic
character of the bulk is practically recovered from the fourth layer and 
beyond. For the Co-Mn/Mn-Al interface, on the HMF side, one observes
a 10$\%$ decrease of the moment at the Mn site with respect to its
bulk value. In the second layer, instead, the moments at the Co sites
are slightly larger than in the bulk. As in the Co-Co/V-Al interface,
they also exhibit a symmetry-breaking, although not so evident. On the
SC side, at the interface, one observes the formation of an induced moment
at the Co site. The presence of a large anti-parallel moment at the Mn site
results in a total moment per layer which is anti-parallel to 
the total magnetization. The moments per layer on the SC side of the 
interface are ordered anti-ferromagnetically along the (001) direction.
This is slightly different than what happens at the Co-Co/V-Al interface,
where interfacial effects cause the first layer (V-Al) and the second 
layer (Co-Mn) to have parallel moments orientation. 

The LSDA results reported in Fig.~\ref{fig:moments} are similar
to the ones obtained by Chadov {\itshape{et al.}}~\cite{ch.gr.11}. 
As a general tendency, the moments at the transition metals
sites seem to be slightly larger in our study than in theirs.
This is partly due to intrinsic reasons, i.e. the existence of
discrepancies between the electron densities calculated in
a full-potential (FP) scheme versus the atomic-sphere 
approximation (ASA), used respectively in this work and in 
Ref.~\onlinecite{ch.gr.11}. The usage a FP scheme is 
expected to be important especially for the layers at the
interface. However, we expect that another reason why we report
larger moments with respect to Ref.~\onlinecite{ch.gr.11} is 
the fact that we use slightly smaller muffin-tin spheres 
to obtain locally-projected quantities. In fact, in ASA one is 
required to use overlapping spheres to cover the full physical space, 
including the interstitial regions between different atoms, while 
in a FP scheme the interstitial
region is treated separately.

In Fig.~\ref{fig:moments} one can also observe the magnetic 
moments obtained in LSDA+U and LSDA+DMFT. The physical picture
obtained in these two approaches is rather similar to the one
obtained in plain LSDA. Including correlation effects leads
to an increase of the magnetic moments at the transition 
metal atoms. The increase is more marked in LSDA+U than in 
LSDA+DMFT and is particularly evident for the Mn and V
sites on the SC side of both interfaces. For example
the moment at the Mn (V) site in LSDA+U is about 2.5 (3.2) 
times larger than in LSDA. In fact, in LSDA+U, the magnitude 
of the V moment in the second layer of the Co-Mn/Mn-Al 
interface is comparable to that of the Co moment in the first
layer.
Another interesting observation is that in LSDA+U the moments
of the Co sites close to the interface become larger than 
those in the bulk, while in LSDA and LSDA+DMFT the situation
is reversed. 

\begin{figure*}[t]
\centering
\includegraphics[width=0.9\linewidth]{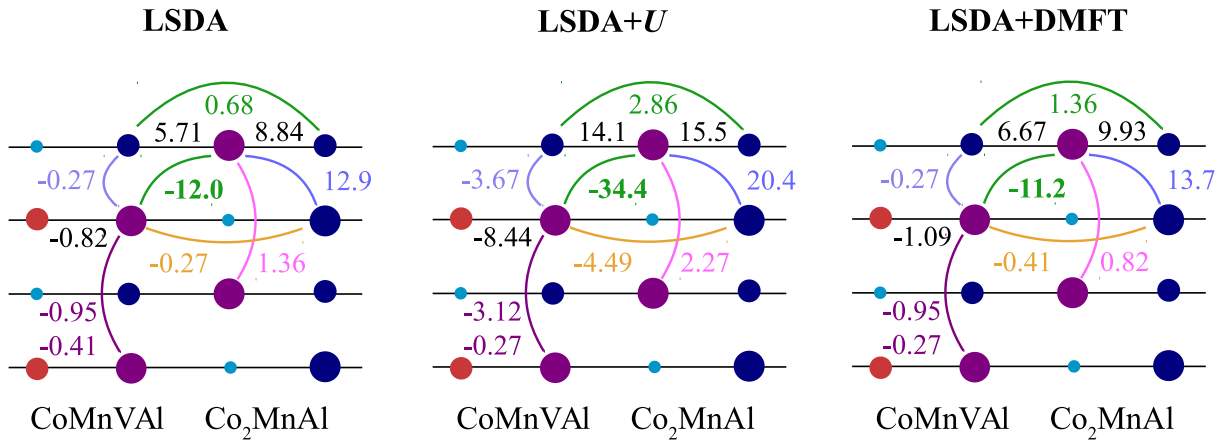}
\caption{(color online) Interatomic exchange parameters $J_{ij}$ in 
meV for the most relevant atomic pairs at 
the interface Co-Mn/Mn-Al. Mn, Co, V and Al atoms are respectively
represented as violet, blue, read and turquoise spheres. Different
exchange parameters are reported with respect to their interatomic
bonds (in the same colours). Bigger circles represent atoms 
belonging to $xz$ plane with $x=0.0$, while small circles represent
atoms belonging to the plane with $x=0.5$. The $z$ axis is perpendicular
to the interface plane as in other plots. Notice that two different 
numbers are reported for the Mn-Mn bond on the side of interface
containing CoMnVAl. These numbers refer to two different directions
which are indistinguishable in a 2-dimensional plane. The upper and 
lower numbers correspond to bonds along the [100] and [010] directions,
respectively.}
\label{fig:j-si}
\end{figure*}


Finally, we have calculated the exchange parameters for the heterostructure. 
We did not consider the Co-Co/V-Al interface because of the negligible
magnetic moments at the SC side. The dominant $J_{ij}$'s at 
the Co-Mn/Mn-Al interface are instead reported in Fig.~\ref{fig:j-si}. 
The most interesting magnetic interactions are the ones across the interface. 
In particular, the largest coupling happens between two Mn atoms at the 
opposite sides of the interface and is antiferromagnetic (green bold number),
which is also reflected by the anti-parallel moment observed in 
Fig.~\ref{fig:moments}. We stress that this strong antiferromagnetic coupling
emerges only at the interface, since in the bulk of Co$_2$MnAl,
Mn ions are only second nearest-neighbours of each other (see Fig.~\ref{fig:j-bulk}).
This interaction is also particularly affected by the inclusion of strong
correlation effects and in LSDA+U becomes even larger than the Co-Mn 
nearest-neighbour coupling that drives the ferromagnetic order of the bulk HMF.
The Co-Mn interaction across the interface (black line in Fig.~\ref{fig:moments})
is ferromagnetic, but weaker than the aforementioned Mn-Mn coupling. Further,
the $J_{ij}$ between Co and Mn on the CoMnVAl side of the interface (purple line)
is antiferromagnetic, but relatively small, even in LSDA+U. Again on the 
CoMnVAl side of the interface, one can see the presence of a small anti-ferromagnetic
coupling between Mn and V (black lines). This coupling increases of an order of 
magnitude in LSDA+U, reaching a strength comparable with the other magnetic interactions.
As discussed for the magnetic moments, LSDA+DMFT results in magnetic couplings that 
are very similar to those obtained in standard LSDA.

Within the current magnetic configuration all the magnetic interactions 
seem satisfied, suggesting that the chosen magnetic order is locally stable.
However, the $J_{ij}$'s extracted by the magnetic force theorem are known to 
depend on the reference state.
Thus, we believe that there are two indications that the current magnetic order
might not be the ground state. First, the Co and Mn moments on the CoMnVAl side
are quite small in magnitude; second, the $J_{ij}$ between them is surprisingly
small. Our interpretation is that this coupling (indicated by a purple line in
Fig.~\ref{fig:j-si}) might actually be ferromagnetic.  However, since it is in
competition with a strong antiferromagnetic Mn-Mn coupling and a strong 
ferromagnetic Co-Mn coupling across the interface, 
the system finds it energetically favourable to suppress these magnetic moments.
To verify this hypothesis we performed additional calculations 
by removing the HMF from the supercell. 
As a matter of fact, we obtain (data not shown) that the corresponding 
magnetic moments are ferromagnetically coupled and become strongly enhanced.
This situation may potentially lead to a non-collinear spin order, which has 
already been suggested for interfaces involving Heuslers alloys~\cite{fe.st.15}. 
A more quantitative analysis of such ordering could be done by performing 
simulations based on atomistic spin dynamics~\cite{et.be.15,eriksson2017atomistic},
but this task is beyond the scope of the present study.

\begin{figure}[t!]
\centering
\includegraphics[width=0.8\linewidth]{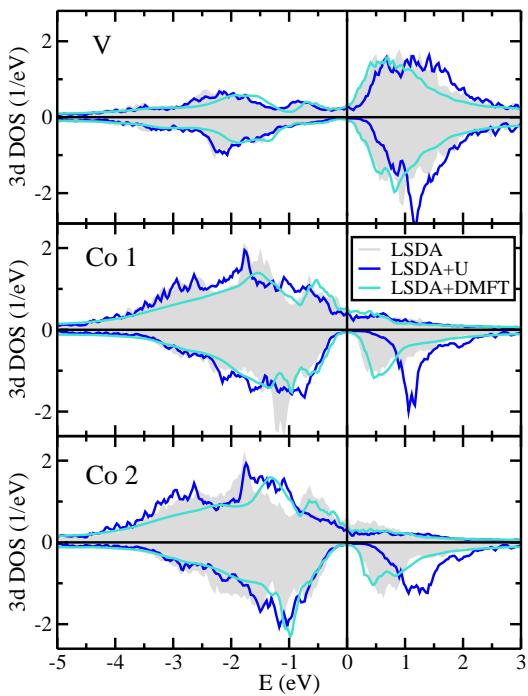}
\caption{(color online) PDOS for the $3d$ states of V and Co (both
types) at the interface Co-Co/V-Al, as obtained in LSDA (gray shade), 
LSDA+U (blue) and LSDA+DMFT (turquoise). V belongs to the SC side of the 
interface, while Co1 and Co2 are located in the HMF side.
Positive and negative values 
correspond to majority and minority spin channels, respectively.
The Fermi level is at zero energy and emphasized with a 
vertical line.}
\label{fig:coco_val_dos}
\end{figure}

\begin{figure}[t!]
\centering
\includegraphics[width=0.8\linewidth]{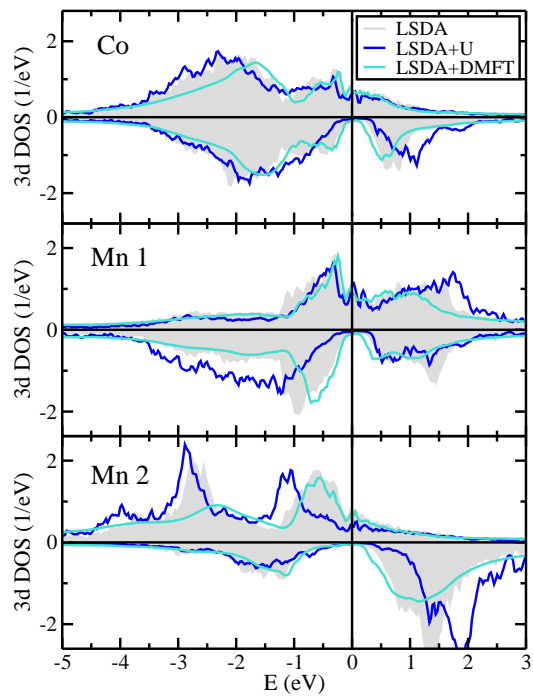}
\caption{(color online) PDOS for the $3d$ states of Co and Mn (both
types) at the interface Co-Mn/Mn-Al, as obtained in LSDA (gray shade), 
LSDA+U (blue) and LSDA+DMFT (turquoise). Co and Mn1 belong to the SC side
of the interface, while Mn2 is located in the HMF side.
Positive and negative values correspond to majority and minority 
spin channels, respectively. Notice that the magnetic moment at the Mn 1 
site is anti-parallel to the global magnetization, therefore there are 
more electrons with minority spin character than majority spin character.
The Fermi level is at zero energy and emphasized with a 
vertical line.}
\label{fig:comn_mnal_dos}
\end{figure}

\subsection{Spectral properties}
We now focus on the half-metallic character
(or lacking of) at the interfaces.
In Fig.~\ref{fig:coco_val_dos}
the PDOS for the $3d$ states of the transition metals 
at the Co-Co/V-Al interface are reported. The minority-spin
band gap is preserved in all computational approaches, but
becomes particularly large in LSDA+U. It is interesting
to see that the corrections induced by the Hubbard terms
are much larger at the Co site with the largest moment
(bottom panel of Fig.~\ref{fig:coco_val_dos}) than at
the Co site with the smallest moment
(middle panel of Fig.~\ref{fig:coco_val_dos}). The 
corresponding self-energies of the Co-$3d$ states
(data not shown) are very
similar to the bulk values of Co$_2$MnAl, with no 
evident signature of NQP states. We can then move 
to the Co-Mn/Mn-Al interface, whose PDOS for all 
relevant states are reported in 
Fig.~\ref{fig:comn_mnal_dos}. 
Also in this case all approaches preserve
the half-metallic character. In LSDA+U one 
observes an increase of the band gap as
well as of the exchange splitting. In LSDA+DMFT,
one observes similar features as in the bulk 
Co$_2$MnAl, even for the Mn site at the SC side
of the interface (``Mn 1'' in Fig.~\ref{fig:comn_mnal_dos}).
Notice that the local magnetization axis at this site is 
different than the global magnetization axis, 
because of the negative magnetic moment 
(see Fig.~\ref{fig:moments}).
For both Mn types at the interface the tail due to the NQP 
states extends closer to the Fermi level than in the bulk HMF,
resulting in a 35\%-reduction of the band gap with respect to
its LDA value.

\begin{figure}[t]
\centering
\includegraphics[width=0.9\linewidth]{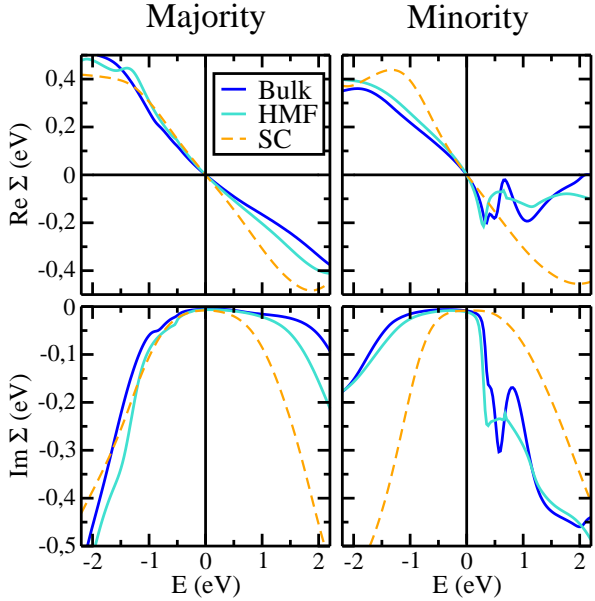}
\caption{(color online) Average self-energy per orbital of the Mn-$3d$ states 
in Co$_2$MnAl and at the HMF side of the Co-Mn/Mn-Al interface, 
separately for majority (left panels) and 
minority (right panels) spin channels. Both real and imaginary 
parts are reported, respectively in the top and bottom panels. The 
Fermi level is at zero energy and emphasized with a vertical line.}
\label{fig:comnmnal_sigma}
\end{figure}

The inspection of the self-energy, reported in Fig.~\ref{fig:comnmnal_sigma},
reveals that the NQP state appears only at the Mn site belonging to the
HMF side of the interface. With respect to the bulk Co$_2$MnAl, the 
NQP state is shifted closer to the Fermi level, which reflects a stronger
reduction of the band gap, mentioned above. Conversely the self-energy
at the Mn site on the SC side of the interface does not show such a feature
and its overall similar to the one for the bulk CoMnVAl (data not shown).
The only difference concerns the absolute intensity, which for the 
majority (minority) spin channel is smaller (bigger) than in the bulk.
Notice that we again refer to the global magnetization axis.

%

\section{Conclusions}
\label{sec:conclusions}
In this work we investigate the correlated band structure and magnetism of the Co$_2$MnAl/CoMnVAl heterostructure, as well as its bulk constituents. Bulk Co$_2$MnAl is shown to be a HMF, whose magnetic moments and exchange couplings depend only mildly on the inclusion of strong electronic correlations (both static and dynamic). In LSDA+U, where the largest corrections are observed, the magnetic moments are increased of about 10\% with respect to their values in LSDA, while the increase of the nearest neighbor exchange coupling is at most of 40\%. No qualitative changes, as e.g. in the sign of the magnetic moments or the interatomic exchange parameters, are observed. LSDA+DMFT simulations clearly show the appearance of NQP states within the minority-spin gap in the band structure. These states are identified to originate mainly from Mn-$3d$ states and are located several hundreds meV above the Fermi level. Therefore, they do not seem to affect the predicted spin polarization very much. Bulk CoMnVAl is shown to be a semi-metal in both LSDA and LSDA+DMFT, while it turns into a proper SC in LSDA+U. The scenarios predicted by these three methods are all compatible with experimental observations and previous literature, pointing to a very limited conductivity. 

The Co$_2$MnAl/CoMnVAl heterostructure is the most interesting system addressed in this work. In LSDA, two distinct interfaces are predicted to have a half-metallic character, namely Co-Co/V-Al and Co-Mn/Mn-Al. This prediction is not changed by including strong correlation effects, both through LSDA+U or LSDA+DMFT. Our LSDA+DMFT calculations predict the appearance of NQP states at the Co-Mn/Mn-Al interface, but this does not affect the spin-polarization at the Fermi energy, similarly to bulk Co$_2$MnAl. The Co-Mn/Mn-Al interface is also interesting for its magnetism. The presence of two Mn atoms which are relatively close to each other leads to a strong antiferromagnetic coupling, which is stabilized by reducing the size of neighboring moments. Although the exchange interactions suggest that the calculated collinear magnetic structure is locally stable, the suppression of some magnetic moments indicates the presence of competing magnetic interactions, which are likely to induce non-collinearity of spins at the interface. The interatomic exchange parameters reported in this work can in principle be used to perform finite-temperature simulations of magnetism through atomistic spin dynamics, which may shed further light on the magnetic ground state and its ordering temperature. Finally, the Co-Co/V-Al interface does not possess any of the intriguing features observed for the Co-Mn/Mn-Al interface. Although this makes it less interesting from a physical point of view, it also makes it more suitable for constructing heterostructures with a high spin-polarization.

\begin{acknowledgments}
We thank Iulia Emilia Brumboiu for the artwork and Johan Sch\"{o}tt for 
performing the analytical continuation. 
Financial support by the DFG Research unit FOR~1348 and by the Knut and Alice Wallenberg Foundation (KAW Projects
No. 2013.0020 and No. 2012.0031) is gratefully acknowledged. The computations were performed on resources provided by the Swedish National Infrastructure for Computing (SNIC) at the National Supercomputer Centre (NSC) at Link\"oping University (Sweden).
\end{acknowledgments}

\appendix

\section{The role of U and J}\label{sec:calculatedU}
The calculations presented in the main text were also repeated 
using Coulomb interaction parameters for Co$_2$MnAl as calculated
in Ref.~\onlinecite{sa.ga.13}. This means that $U$ ($J$) was set 
to 3.23 (0.6) eV and 3.40 (0.7) eV for respectively Mn and Co.
In principle, there are several reasons why using directly those values
does not represent a correct procedure in our computational scheme.
\begin{figure}[b]
\centering
\includegraphics[width=0.8\linewidth]{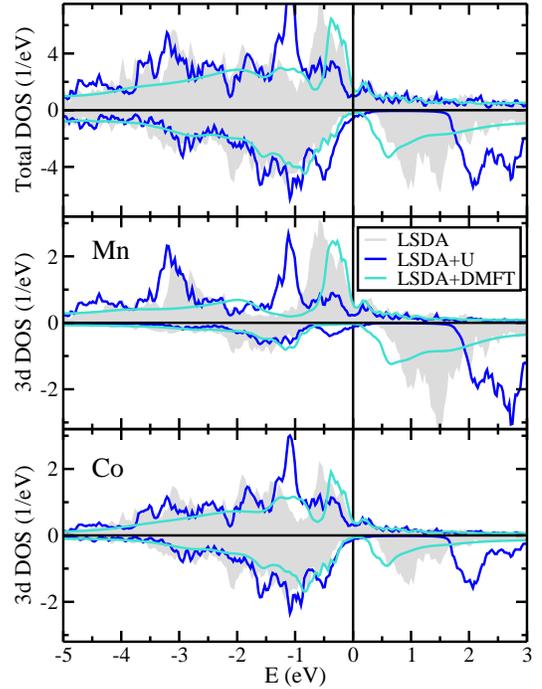}
\caption{(color online) As Fig.~\ref{co2mnal_dos}, but for different values
of $U$ and $J$.}
\label{fig:dos_summary_HM_U3}
\end{figure}
First, these values were obtained
for a localized basis set (maximally localized Wannier functions) that is
different
from the one used here (atomic-like functions evaluated at the linearization
energy of the LMTOs; see Ref.~\onlinecite{gr.ma.07} for more details). 
Secondly, neither SPTF nor LSDA+U offer a 
proper description of the intra-$3d$ screening (as a matter of fact 
in LSDA+U there is no screening at all, being a single-particle 
approximation at the Hartree-Fock level). Therefore using calculated 
values from cRPA would lead to a drastic overestimation of correlation
effects beyond DFT.

\begin{figure}[t]
\centering
\includegraphics[width=0.9\linewidth]{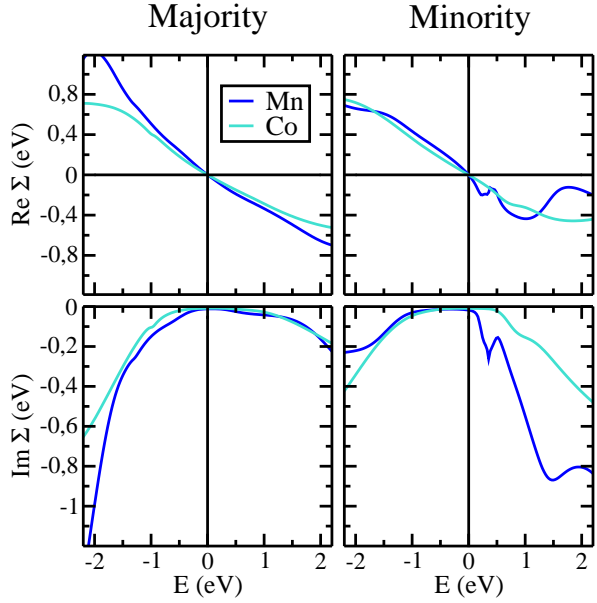}
\caption{(color online) As Fig.~\ref{co2mnal_sigma} but for different 
values of $U$ and $J$.}
\label{co2mnal_sigma_U3}
\end{figure}

\begin{figure}[b]
\centering
\includegraphics[width=0.8\linewidth]{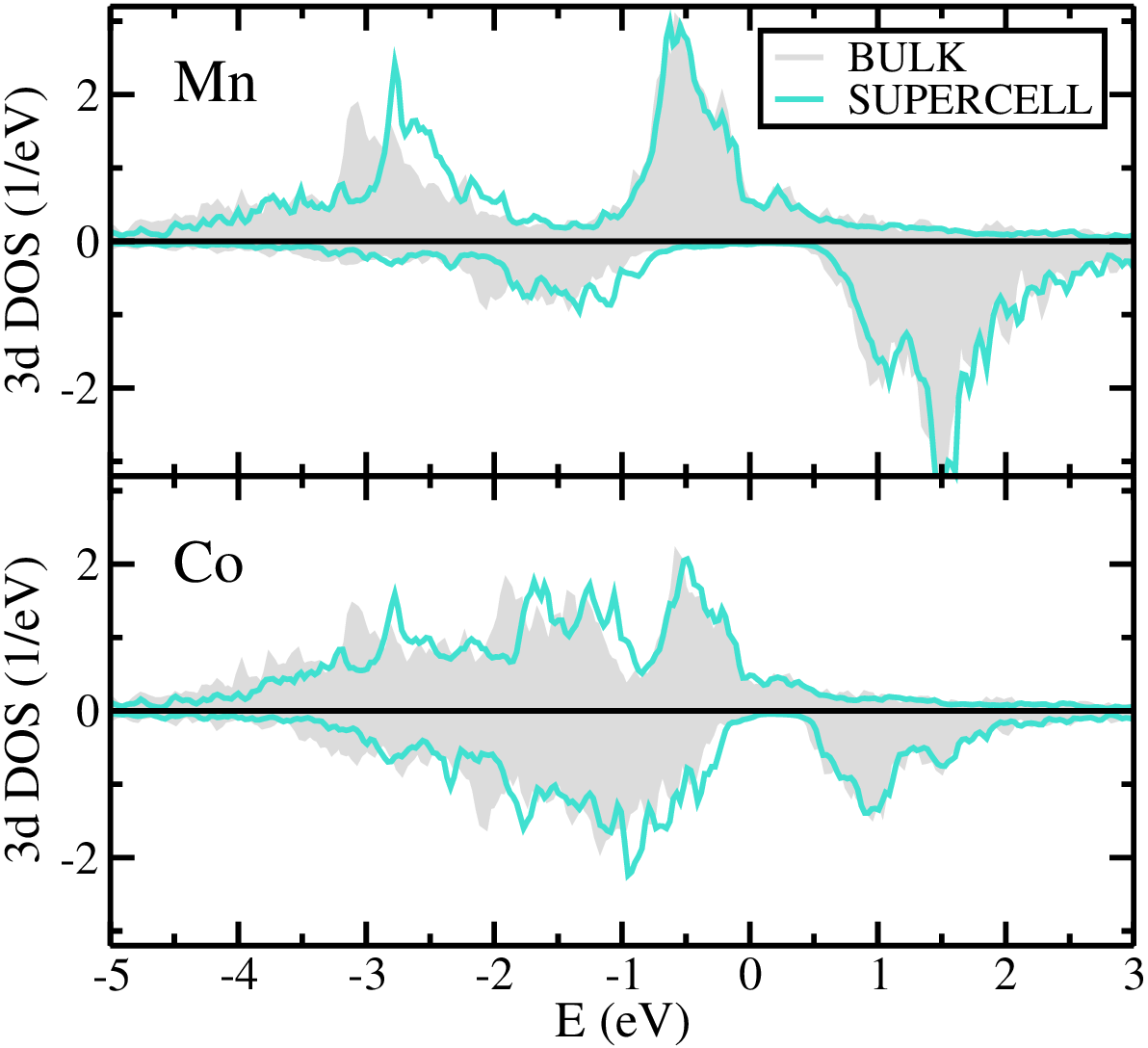}
\caption{(color online) PDOS for the $3d$ states of Mn and Co,
in bulk Co$_2$MnAl (gray shade) and in the innermost layer (turquoise)
of the half-metallic
side of the heterostructure. Calculations were done in LSDA.
Positive and negative values correspond to majority and minority 
spin channels, respectively. The Fermi level is at zero energy 
and emphasized with a vertical line.}
\label{fig:pseudobulk_dos_HM}
\end{figure}

In any case, the results obtained from these additional calculations offer a 
physical picture close to the one discussed in Sec.~\ref{sec:co2mnal}.
We can illustrate these points with the help of 
Fig.~\ref{fig:dos_summary_HM_U3}, where DOS and PDOS for all the relevant
states are reported. As a consequence of a larger value of $U$, LSDA+U
predicts a larger gap in the minority spin channel and the formation of 
a pseudogap in the majority spin channel. For even larger values of $U$, 
this system is likely to turn into an insulator. In LSDA+DMFT, instead,
one can see that the signature of NQP states in the minority spin 
channel is much more evident, as a tail of states propagates from the 
conduction band towards the Fermi level. Even for this overestimated 
value of $U$, however, the half-metallic character is not broken,
which confirms the whole analysis presented in the main text. Also 
concerning the self-energy, which is reported in 
Fig.~\ref{co2mnal_sigma_U3},
we do not observe qualitative differences, but just a overall increase.
The peak corresponding to the NQP states retains the same magnitude
but shifts slightly closer to the Fermi energy.

\begin{figure}[b!]
\centering
\includegraphics[width=0.8\linewidth]{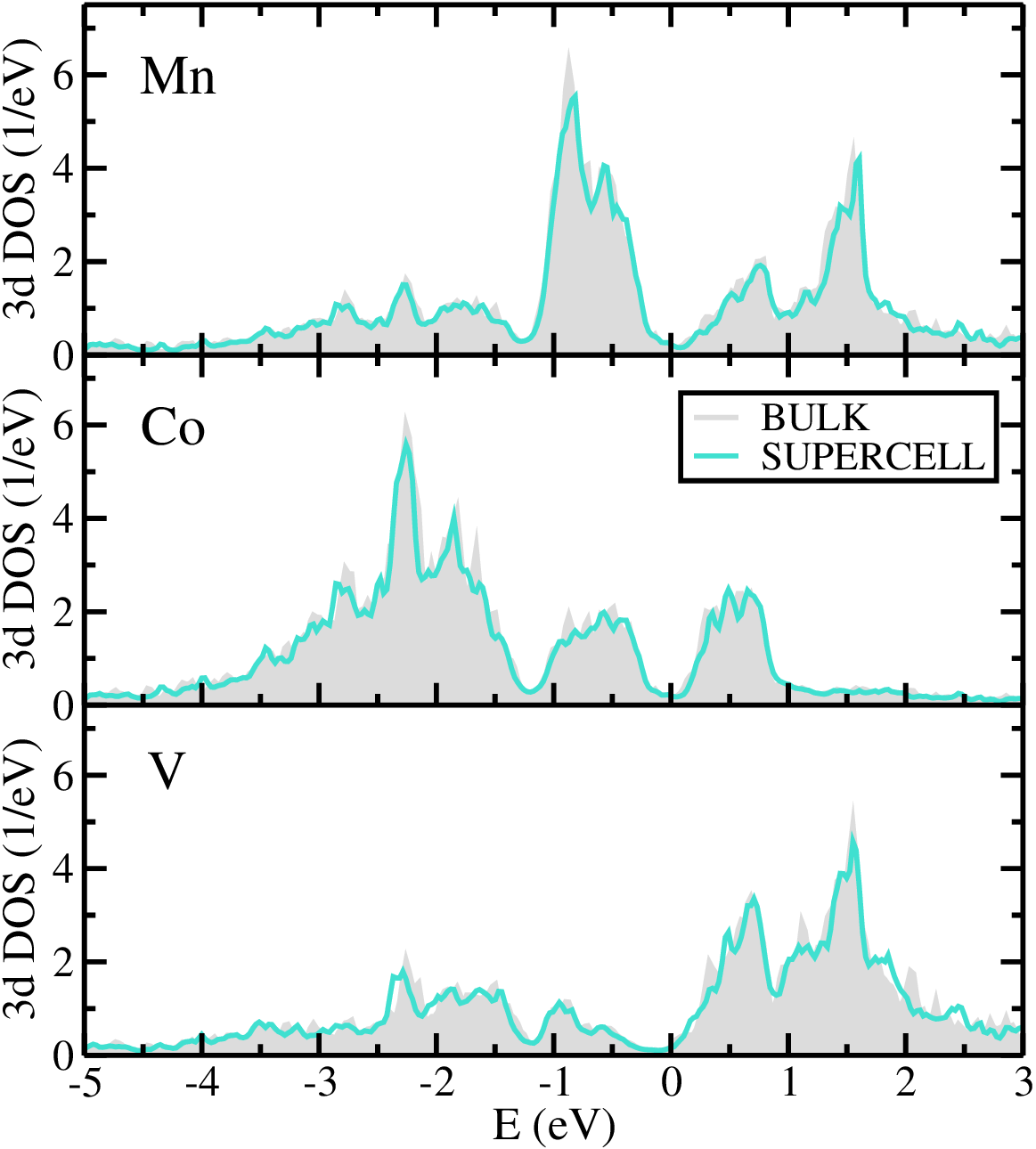}
\caption{(color online) PDOS for the $3d$ states of Mn, Co and V,
in bulk CoMnVAl (gray shade) and in the innermost layer (turquoise)
of the semiconducting
side of the heterostructure. Calculations were done in LSDA.
Positive and negative values correspond to majority and minority 
spin channels, respectively. The Fermi level is at zero energy 
and emphasized with a vertical line.}
\label{fig:pseudobulk_dos_SC}
\end{figure}

We also performed similar calculations for CoMnVAl, by using Coulomb
interaction parameters calculated for Co$_2$MnAl and Mn$_2$VAl in
Ref.~\onlinecite{sa.ga.13}. This means that $U$ ($J$) was set 
to 3.23 (0.6) eV, 3.40 (0.7) eV and 3.06 (0.55) for respectively 
Mn, Co and V. These parameters are not ideal, not only for 
the reasons mentioned above, but also since Co$_2$MnAl and Mn$_2$VAl
are both
HMFs, while CoMnVAl is a SC. The physical picture emerging from 
LSDA+U calculations is similar to the one already discussed in 
in Sec.~\ref{sec:comnval}. DOS and PDOS (data not shown) are 
similar to the LSDA+U curves reported in Fig.~\ref{dossc}, but 
characterized by a slightly larger gap. The LSDA+DMFT calculations,
instead, show that for this large value of $U$, CoMnVAl becomes
a metal. This does not seem a realistic effect, but 
is likely to be due to the perturbative nature of SPTF, which
is not able to treat such large values of the Coulomb interaction
parameters in absence of magnetism. This problem has been 
extensively discussed in Ref.~\onlinecite{di.mi.09}.

\section{Details of the supercell}\label{sec:supercell}
In this study we used a supercell of 64 atoms, shown
in Fig.~\ref{fig:interface}. The innermost layers of
each side (component) were found to be in reasonable 
agreement with their corresponding bulk systems. 
Magnetic moments as obtained in LSDA
are reported in Table~\ref{tab:pseudobulk mag}.
As one can see, the properties of bulk CoMnVAl are 
reproduced quite well, with all magnetic moments
disappearing in the layers of the supercell which 
are farther from the interfaces. Concerning Co$_2$MnAl, 
instead, there is some discrepancy for the magnetic
moment at the Mn site, which in the supercell is 
about 4\% larger than in the bulk. The moments at 
the Co sites, conversely, are about 3\% smaller in
the supercell than in the bulk. 
Similar conclusions are reached by looking at
the PDOSs for the electronic states of interest,
which are reported in 
Figs.~\ref{fig:pseudobulk_dos_HM} 
and~\ref{fig:pseudobulk_dos_SC}. The PDOS of 
Co$_2$MnAl in the bulk and in the supercell 
show some differences, but those are mainly located
far from the Fermi level and therefore do not change
our qualitative analysis of the properties at the
interface. Finally, projections over real spherical
harmonics (not shown) illustrate that in the innermost
layers the symmetry of the bulk states, which is broken
at the interface, is completely recovered. Reproducing
the exact crystal field splitting observed in the 
bulk, however, requires a few more layers of 
Co$_2$MnAl.

\begin{table}[b]
\begin{tabular}{cccccc}
\hline
     &  $\mu_{Mn}$ &  $\mu_{Co}$ & $\mu_{Al}$ \\
\hline
Co$_2$MnAl    & 2.48 (2.59) & 0.81 (0.78) & -0.03 (-0.03) \\
CoMnVAl       & 0.00 (0.00) & 0.00 (0.00) & 0.00 (0.00) \\
\hline
\end{tabular}
\caption{Site-projected and total magnetic moments for bulk Co$_2$MnAl
compared to the corresponding values (within parentheses) in the innermost
layer of the HMF side of the interface. The values are expressed in 
$\mu_{B}$ and are obtained in LSDA.}
\label{tab:pseudobulk mag}
\end{table}

\end{document}